\documentclass[
    ,final            
  ]
  {aipproc}

\layoutstyle{6x9}

\def\DZ{D0 }

\def\vzRF{4.47}

\def\vzRFstat{0.64}

\def\vzRFsystu{0.73}
\def\vzRFsystd{0.72}

\def\vzRFnsigma{4.6}

\def\vznlo{4.4}   
\def\vznloe{0.3}  

\def\vzresult{$\sigma(WW+WZ)=\vzRF\pm\vzRFstat$~(stat)
  $^{+\vzRFsystu}_{-\vzRFsystd}$~(syst)~pb}

\begin{document}

\title{Searches for the Higgs boson at the Tevatron}


\author{Yuri Oksuzian
On behalf of the CDF and D0 collaborations}{
  address={University of Virginia, Charlottesville, Virginia 22904, USA}
}

\begin{abstract}
We present the results from the Tevatron on the direct searches for
Standard Model Higgs boson produced in $p\bar{p}$ collisions at a
center of mass energy of 1.96 TeV, using the data corresponding to the
integrated luminosity of 10fb$^{-1}$. The searches are performed in the
Higgs boson mass range from 100 to 200 GeV/c$^{2}$. The dominant production
channels, $H\rightarrow b\bar{b}$ and $H \rightarrow WW$, are combined with all the secondary
channels and significant analysis improvements have been implemented
to maximize the search sensitivity. We observe a significant excess of
data events compared to background predictions with the local
significance of 3.0 standard deviations. The global
significance for such an excess anywhere in the full mass range
investigated is approximately 2.5 standard deviations. 
\end{abstract}

\maketitle


\section{Introduction}

The Higgs boson is a crucial element of the standard model (SM) of elementary particles
and interactions. Within the SM, vector boson masses arise from the spontaneous breaking of
electroweak symmetry due to the existence of the Higgs particle. The winter results from the LHC and
the Tevatron experiments have excluded wide regions of the possible
Higgs mass ranges. The most interesting region to search for the Higgs is the mass range between 115 and 127 GeV/$c^2$
where the both the ATLAS and the CMS experiments have found some excesses~\cite{atlas,cms}. The Tevatron experiments can contribute to the understanding
of this region by analyzing the data collected through the years of
2001-2011.

\section{Higgs Search Channels at the Tevatron}

\subsection{Low Mass Channels}

The SM Higgs boson $H$ is predicted to be
produced in association with a $W$ or $Z$ boson at the Fermilab Tevatron 
$p{\bar{p}}$ collider and its dominant decay mode is predicted to be into a bottom-antibottom quark pair ($b{\bar{b}}$), 
if its mass $m_H$ is less than 135~GeV/$c^2$~(low Higgs mass region). The searches use the complete Tevatron data sample
of $p{\bar{p}}$ collisions at a center of mass energy of 1.96 TeV collected
by the CDF and D0 detectors at the Fermilab Tevatron, with an integrated
luminosity of 9.45 fb$^{-1}$~--~9.7~fb$^{-1}$. The CDF and D0 detectors are multipurpose solenoidal spectrometers surrounded by 
hermetic calorimeters and muon detectors and are designed to study the products
of 1.96 TeV proton-antiproton
collisions~\cite{cdfdetector,d0detector}.

The online event selections (triggers) rely on fast reconstruction
of combinations of high-$p_T$ lepton candidates, jets, and
$\mbox{$\not\!\!E_T$}$. Event selections are similar in the CDF and D0
analyses, consisting typically of a preselection based on event topology and kinematics,
and a subsequent selection using
$b$-tagging.  Each channel is divided into exclusive sub-channels according to various
lepton, jet multiplicity, and $b$-tagging characterization criteria aimed at
grouping events with similar signal-to-background ratio and so optimize
the overall sensitivity.

Due to the importance of $b$-tagging,
both collaborations have developed multivariate approaches to maximize the performance
of the $b$-tagging algorithms.  
A boosted decision tree algorithm is used in the D0 analysis, which builds and improves
upon the previous neural network $b$-tagger~\cite{Abazov:2010ab},
giving an identification efficiency of
$\approx 80\%$ for $b$ jets with a mis-identification rate of
$\approx 10\%$.  The CDF $b$-tagging algorithm has been recently augmented
with an MVA~\cite{tagging}, providing a $b$-tagging efficiency
of $\approx 70$\% and a mis-identification rate of $\approx 5$\%.

In $H\rightarrow b\bar{b}$ final states, the single most sensitive observable to distinguish between the Higgs
signal and various types of background is the invariant mass of dijet
system, $m_{jj}$, which approximately accounts for 75\% of analysis
sensitivity. In all low mass Higgs searches at Tevatron, we include
additional variables through the multivariate analysis techniques. Dedicated studies have been performed to improve the search
sensitivity through the improvements in dijet mass resolution, lepton
identification algorithm, $b$-tagging, multijet background suppression
and modeling, final discriminant optimization. The detailed
information on low mass Higgs channels is present in Ref. ~\cite{cdfwh2012,cdfzh2012,cdfzhll2012,dzwh2012,dzzh2012,dzzhll2012}

To validate our background modeling and search methods, we
perform a search for SM diboson production in the same final 
states used for the SM $H\rightarrow b\bar{b}$ searches. The 
data sample, reconstruction, process modeling, uncertainties, and 
sub-channel divisions are identical to those of the SM Higgs boson 
search.  The measured cross section for $WZ$ and $ZZ$
production is \vzresult~\cite{mor12tevdibosons}. This is consistent  with SM prediction
of $\sigma(WW+WZ)=\vznlo\pm\vznloe$~pb~\cite{dibo} and corresponds to a significance of \vzRFnsigma~standard deviations above the background-only hypothesis.

\subsection{Other Complimentary Channels}

Even though $H\rightarrow b\bar{b}$ final states are the most sensitive channels at the
Tevatron below 135 GeV/c$^2$, in the final combination we consider all the complimentary
channels to improve the Higgs search sensitivity. The complete list of channels that goes into the Higgs Tevatron
combination is given in Ref.~\cite{higgscombo}. One of the channels
that needs to be mentioned is $H \rightarrow WW$. Being the most
sensitive channel for the high mass Higgs region, it has significant
contribution to the low mass region as well. For the $H \rightarrow WW$ analyses, signal events are characterized
by large  $\mbox{$\not\!\!E_T$}$ and two opposite-signed, isolated leptons.  The presence of neutrinos in the
final state prevents the accurate reconstruction of the candidate
Higgs boson mass. The most sensitive variable for Higgs signal is the
opening angle,
$\Delta R$, between the outgoing leptons. Both CDF and D0  include
additional event properties and their correlations through multi
variate algorithms. CDF uses neural network outputs, including
likelihoods constructed from calculated matrix element 
probabilities and D0 uses boosted decision trees outputs.


\section{Tevatron Combination}

In the Tevatron combination, we combine all the major low mass Higgs channels
with the complimentary final state searches from CDF and D0. In this
section we report the results presented at CIPANP
conference, which are based on analyses presented for the Winter conferences
and described in more detailes in Ref.~\cite{higgscombo}.

To determine the estimates of the interest like the upper limits on SM Higgs production at 95\% C.L. and to gain confidence that the final result does not depend on the
details of the statistical formulation, we perform two types of combinations:
Bayesian approach where the nuisance
parameters are integrated out to determine posterior probabilities;
and  Modified Frequentist approach where the minimum of the likelihood is used to determine
the nuisance parameters. Both approaches yield limits on the Higgs boson
production rate that agree within 10\% at each
value of $m_H$, and within 1\% on average. Systematic uncertainties enter on the
predicted number of signal and background events as well
as on the distribution of the discriminants in
each analysis (``shape uncertainties'').

\begin{figure}
  \includegraphics[height=.3\textheight]{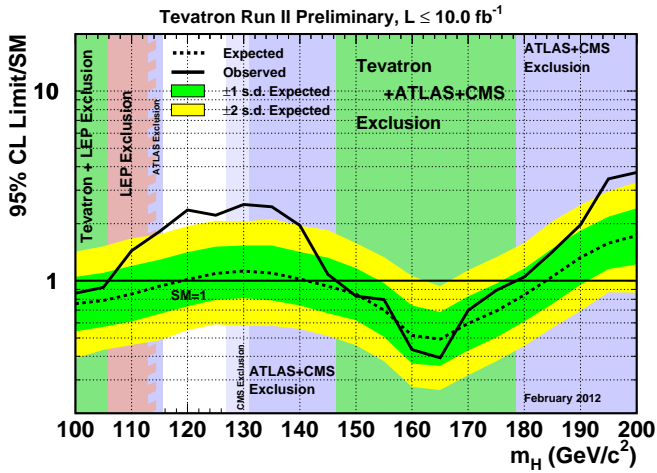}
  \includegraphics[height=.277\textheight]{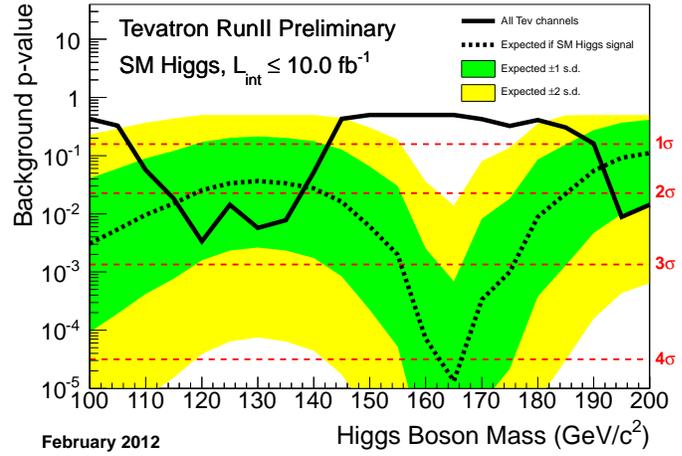}
  \caption{
 \label{fig:TevAll} 
Final Tevatron combination for the winter conferences: (Left) The observed 95\% credibility level upper limits on SM Higgs
boson production as a function of Higgs boson mass.  The dashed line
indicates the median expected value in the absence of a
signals. (Right) The $p$-value as a function of $m_H$ under the
background-only hypothesis.  The associated dark and light-shaded
bands indicate the 1 s.d.~and 2 s.d.~fluctuations of possible
experimental outcomes.}
\end{figure}

The 95\% C.L. limits on Higgs production are shown in Fig.~\ref{fig:TevAll}, along
with the significance of the excess in the data over the background
prediction, assuming a signal is truly absent.  The regions of Higgs
boson masses excluded at the 95\% C.L. are $100<m_H<106$~GeV/$c^{2}$
and $147<m_{H}<179$~GeV/$c^{2}$.  The expected exclusion regions are
$100<m_H<119$~GeV/$c^{2}$ and $141<m_{H}<184$~GeV/$c^{2}$.  There is
an excess of data events with respect to the background estimation in
the mass range 115<$m_H$<135 GeV/$c^2$. The observed $p$-value as a function of $m_H$ exhibits
a broad minimum, and the maximum local significance corresponds to 2.7
standard deviations at $m_H=120$~GeV/$c^2$. Correcting for the
Look-Elsewhere Effect (LEE), which accounts for the 
possibility of a background fluctuation affecting the local $p$-value anywhere 
in the search region,  yields a global significance of 2.2 standard deviations.

\section{Updated results}

We have recently updated and combined the results in $H\rightarrow b\bar{b}$ final
states at CDF and D0~\cite{TevBBprl}. An observation of this process would support
the SM prediction that the mechanism for electroweak symmetry breaking, which
gives mass to the weak vector bosons, is also the source of fermionic mass in the
quark sector. 

\begin{figure}
  \includegraphics[height=.3\textheight]{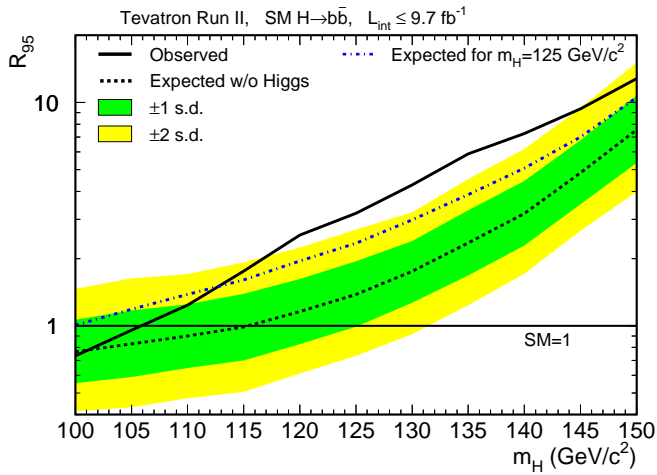}
  \includegraphics[height=.277\textheight]{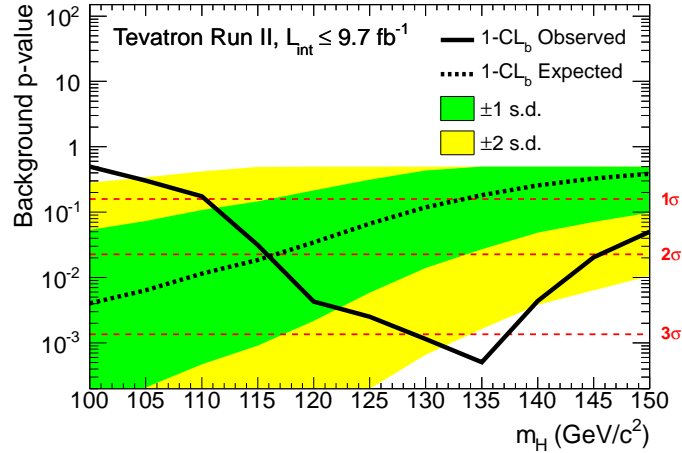}
  \caption{
 \label{fig:TevBB} 
Updated Tevatron combination for $H\rightarrow b\bar{b}$ channels: (Left) The observed 95\% credibility level upper limits on SM Higgs
boson production as a function of Higgs boson mass.  The dashed line
indicates the median expected value in the absence of a
signals. (Right) The $p$-value as a function of $m_H$ under the
background-only hypothesis.  The associated dark and light-shaded
bands indicate the 1 s.d.~and 2 s.d.~fluctuations of possible
experimental outcomes.}
\end{figure}

The broad observed excess in the low mass range, shown on Fig.~\ref{fig:TevBB}, results in a minimum $p$-value of 3.3 standard deviations
away from the background-only hypothesis at a Higgs mass of $m_H$ =
135 GeV/c$^2$. The global $p$-value is 3.1 standard deviation.
We interpret this result as evidence for the presence of a particle that is 
produced in association with a $W$ or $Z$ boson and decays to a bottom-antibottom 
quark pair.  The excess seen in the data is most significant in the mass range 
between 120 and 135~GeV/$c^2$, and is consistent with production of the SM Higgs 
boson.

The updated Tevatron combination~\cite{higgscomboNew} across all channels on CDF and D0 yields the local(global) significance for 
such an excess of 2.5(3.0) standard deviations.

\section{Conclusions}
We combine all available CDF and D0 results on SM Higgs boson searches. A broad excess is observed in data with respect 
to the background estimation, corresponding to a 2.5 standard deviations. Considering only
the $H\rightarrow b\bar{b}$ final state searches yields an excess, corresponding
to a 3.1 standard deviations.  The excess is observed to be consistent with SM Higgs
boson production.

\begin{theacknowledgments}
The author thanks the CIPANP conference organizers, conveners,
colleagues from CDF and D0 experiments, and acknowledges the support
from DOE and visiting scholar award from URA.
\end{theacknowledgments}



\bibliographystyle{aipproc}   


\IfFileExists{\jobname.bbl}{}
 {\typeout{}
  \typeout{******************************************}
  \typeout{** Please run "bibtex \jobname" to optain}
  \typeout{** the bibliography and then re-run LaTeX}
  \typeout{** twice to fix the references!}
  \typeout{******************************************}
  \typeout{}
 }


\end{document}